\documentstyle[sprocl]{article}

\def\Journal#1#2#3#4{{#1} {\bf #2}, #3 (#4)}

\def\NPA{{\em Nucl. Phys.} A}
\def\NPB{{\em Nucl. Phys.} B}

\def\PRD{{\em Phys. Rev.} D}

\newcommand{\hal}[1]{{#1 \over 2}}
\newcommand{\ket}[1]{| #1 \rangle}
\newcommand{\bra}[1]{\langle #1 |}

\newcommand{\pslash}{{p\hspace{-5pt}/}}
\newcommand{\delslash}{\partial \hspace{-6pt}/}
\begin{document}

\title{SUPPRESSION OF $\pi NN^*$ COUPLING AND CHIRAL SYMMETRY}

\author{D. Jido, Y. Nemoto, M. Oka}

\address{Department of Physics, Tokyo Institute of Technology \\
Meguro, Tokyo 152 Japan \\
E-mail: jido@th.phys.titech.ac.jp}

\author{A. Hosaka}

\address{Numazu College of Technology \\
3600 Ooka, Numazu 410 Japan}

% \begin{center}
% {\large Suppression of $\pi NN^*$ Coupling and Chiral 
%     Symmetry\footnote{Talk given by D.J. at APCTP workshop
%   ``Properties of Hadrons in Matter'', Seoul, Korea in October (1997)}\\ }
% \vspace*{0.5cm}
%  { D. Jido \footnote{e-mail address: jido@th.phys.titech.ac.jp}, 
%   Y. Nemoto  and M. Oka \\
%  Department of Physics, Tokyo Institute of Technology\\
%  Meguro, Tokyo 152  Japan}
% 
% \vspace*{0.5cm}
% {A. Hosaka \\
% Numazu College of Technology \\
% 3600 Ooka, Numazu 410 Japan}
% \end{center}
% \vspace*{0.3cm}

\maketitle
\abstracts{ Meson-baryon couplings between positive and negative parity 
nucleons are investigated using two point correlation functions in the 
soft meson and the chiral limit.  We find that the $\pi NN^*$ coupling 
vanishes due to chiral symmetry, while the $\eta NN^*$ coupling 
remains finite.  We perform an analysis based on the algebraic method 
for $SU(2)$ chiral symmetry, and find that nucleon axial charges play 
an essential role for vanishing coupling constants.  For an 
illustration, we construct sigma models with $N$ and $N^*$, where we 
discuss their properties from the view point of the chiral symmetry.  
We also discuss $N$ and $N^*$ in the viewpoint of chiral 
partner.}

\section{Introduction}
Chiral symmetry and its spontaneous breakdown are very important in 
the low energy QCD. Current algebra and low energy theorems have been successfully 
applied to hadronic phenomena such as meson-meson scatterings and 
photoproductions of mesons~\cite{ad}.  
Recent interest in the chiral symmetry is largely related to 
the study of 
the restoration of the chiral symmetry at finite temperature and 
density~\cite{hk}.  
One of the implications there
is the appearance of degenerate particles 
with opposite parities when the chiral symmetry is restored.  
For the meson sector they would be, for example, 
$(\sigma , \vec \pi)$ and $(\vec \rho, \vec a_{1})$.  
In contrast, the role of the chiral symmetry is 
not fully explored in the baryon sector.  
Possible candidates for such would-be degenerate 
particles are $N(939)$ and $N(1535)$.  
However, the fate of these particles towards chiral symmetry 
restoration is not clear.  

Recently we have studied the masses of negative parity spin $\hal{1}$ 
baryons ($B_{-}$) with flavor octet and singlet representations using 
the QCD sum rule approach~\cite{jko,jo}.  In order to calculate the 
mass of a hadron in the QCD sum rule, we usually use a correlation 
function of an interpolating field of the hadron.  First we note that 
the interpolating field $J$ for baryons can couple both with positive 
and negative parity baryons:
\begin{eqnarray}
     \bra{0} J(x) \ket{B_{+}} & = & \lambda_{B_{+}} u_{B_{+}}(x) \ ,
         \label{constN} \\
     \bra{0} J(x) \ket{B_{-}} & = & i \gamma_{5} \lambda_{B_{-}} 
     u_{B_{-}}(x) \ . 
         \label{constN*}
\end{eqnarray}
Therefore, the contributions of positive and negative parity baryons 
are mixed in the correlation function for the positive parity baryons, 
and we need to separate them from the correlation function to 
investigate the properties of the $B_{-}$ in the QCD sum rule.
In order to achieve this,
we proposed a formulation to extract the $B_-$ masses~\cite{jko},
and found that the quark condensates play an essential role for 
the mass splitting between $N$ and $N^*$.  (In this paper $N^*$ 
denotes a nucleon with negative parity, e.g.\ $N(1535)$.)  When the 
chiral symmetry is restored, where the chiral order parameter $\langle 
\bar{q} q \rangle$ goes to zero, we have found that $N^*$ is 
degenerate with the ground state nucleon $N$ and tends to be a 
massless particle.  
The details are explained in our 
papers~\cite{jko,jo}.

The motivation of this work is to investigate the $\pi NN^*$ coupling 
in a similar approach.  Experimentally the decay modes of $N(1535)$ are 
well known, and using the observed decay widths~\cite{PDG}, we find 
two major coupling constants, $g_{\pi NN^*} \sim 0.7$ and $g_{\eta 
NN^*} \sim 2$.  Here we realize that the value of $g_{\pi NN^*} \sim 
0.7$ is much smaller than the typical values of pion-baryon coupling 
constants, e.g.  $g_{\pi NN} \sim 13$.  The purpose of the present 
paper is to show that the chiral property of the nucleon plays a 
crucial role in suppressing $g_{\pi NN^*}$.

This paper is organized in the following way.  In Sec.2 the $\pi NN^*$ 
coupling is investigated based on the QCD sum rule using a two point 
correlation function between the vacuum and a soft pion state.  Due to 
the transformation properties for the interpolating field for $N$ and 
$N^*$, the $\pi NN^*$ coupling vanishes in the soft pion and chiral 
limit.  Extending this argument to the case of $\eta$, we see that the 
$\eta NN^*$ coupling need not disappear.  In Sec.3 in order to 
illustrate this situation, we construct two types of sigma models with $N$ and 
$N^*$, which are different in the 
assignments of the chiral property of $N^*$. 
We see that the chiral assignment plays an important role for the $\pi NN^*$ 
coupling. In the naive assignment, we obtain the same result as the 
QCD sum rule approach, while in another assignment the $\pi NN^*$ 
coupling need not vanish.  A conclusion is given in Sec.4.

\section{$\pi NN^{*}$ in correlation function analysis \protect{\cite{jho2}}}
For the investigation of the $\pi NN^{*}$ coupling we follow the 
method of Shiomi and Hatsuda~\cite{sh}, who have calculated $g_{\pi 
NN}$ coupling in the QCD sum rule using a two point correlation 
function between the vacuum and a soft pion state.  In the soft pion 
limit $(q^{\mu}\to 0)$, the correlation function is given by:
\begin{eqnarray}
   \Pi^{\pi}(p) &=& i \int d^4 x \, e^{ip\cdot x}
        \bra{0} TJ(x;s) \bar{J}(0;t) \ket{\pi (q=0)} \nonumber \\
        &=&
        i  (\Pi_{0}^{\pi}(p^{2})\gamma_{5}  + \Pi_{1}^{\pi}(p^{2}) 
            \pslash\gamma_{5} ) \, .
   \label{eq:cor}
\end{eqnarray}
The nucleon interpolating field $J(x;t)$ is defined 
later in eq.(~\ref{eq:nucur}).  Here the Lorentz structures of the two 
terms of (\ref{eq:cor}) are determined by the total parity of the two 
point function.  In this paper, we consider the behavior of this 
correlation function (\ref{eq:cor}) at the kinematical point where the 
four momenta of $N$ and $N^*$ are both fixed at $p_{\mu}$ and in the 
soft pion limit $q \to 0$.  In this choice, symmetry structure of the 
correlation function is best studied, which is our main interest in 
the present paper.  As it was mentioned before, in order to 
investigate the contribution of the negative parity baryon, we have 
to project it out from the correlation function.

In this paper we use the following interpolating field for the nucleon:
\begin{eqnarray}
   J(x;t) & = & \varepsilon^{abc} [(u_{a}(x)Cd_{b}(x))
   \gamma_{5} u_{c}(x) + t (u_{a}(x) C \gamma_{5} d_{b}(x))
   u_{c}(x)] \, ,
        \label{eq:nucur}
\end{eqnarray}
where $a$, $b$ and $c$ are color indices, and $C = i \gamma_{2} 
\gamma_{0}$ is the charge conjugation matrix.  This is the most general 
form without derivatives with three quarks.  The parameter $t$ 
controls the coupling strength of $J$ to various nucleons.  For 
instance, it is known that the Ioffe's current $J(x;t=-1)$~\cite{i} 
couples strongly to the positive parity nucleon~\cite{ept}, while it 
was found that $J(x;t=0.8)$ is optimal for the negative parity 
nucleon~\cite{jko}.  In this way, the interpolating field 
(\ref{eq:nucur}) has been successfully used in the QCD sum rule to 
study both positive and negative parity nucleons~\cite{jko}.  Later we 
will see that the parameter $t$ also specifies the chiral structure of 
$J(x;t)$, which determines the property of meson-nucleon couplings.

First let us investigate the phenomenological side of the correlation 
function.  In general, a meson-nucleon coupling is pseudoscalar type 
if the initial ($N_{1}$) and final ($N_{2}$) nucleons have the same 
parity, while it is scalar type if $N_{1}$ and $N_{2}$ have the 
opposite parity.  Hence the interaction Lagrangians are written as
\begin{equation}
   L_{\rm int} = \left \{
      \begin{array}{c}
                                g_{\pi N_{1}N_{2}} \bar{N}_{1} i \gamma_{5} \tau^{a} 
                                \pi_{a} N_{2} + {\rm (h.c.)} \hspace{1cm} {\textstyle 
                                \rm for\ the\ same\ parity} \\
                g_{\pi N_{1}N_{2}} \bar{N}_{1} \tau^{a} \pi_{a} N_{2} + {\rm 
                (h.c.)} \hspace{1cm} {\textstyle \rm for\ opposite\ parity} 
                \end{array} \right.  \ .
\end{equation}
Using these Lagrangians we find that $\Pi_{0}^{\pi}(p^{2})$ includes 
both the diagonal couplings (e.g.  $N$-$N$, $N^{*}$-$N^{*}$ etc.\ ) 
and the off-diagonal couplings ( e.g.  $N$-$N^{*}$, $N^*$-$N^{**}$ 
etc.\ ), while $\Pi_{1}^{\pi}(p^{2})$ has only the off-diagonal 
couplings.  The part of $\pi NN^*$ coupling in the off-diagonal term
$\Pi^{\pi}_{1}(p)$ is given by
\begin{equation}
   \label{eq:pNN*}
    g_{\pi NN^*} \lambda_{N}(t) \lambda_{N^{*}}(s)
    \left[{p^{2} + m_{N}m_{N^{*}} \over (p^{2} - m_{N}^{2}) (p^{2} -
    m_{N^{*}}^2)} + { \pslash (m_{N} + m_{N^{*}}) \over (p^{2} -
    m_{N}^{2}) (p^{2} - m_{N^{*}}^2)}\right] i \gamma_{5} \, 
\end{equation}
if $N$ and $N^{*}$ have opposite parities.  If $N$ and $N^{*}$ carry 
the same parity, the mass $m_{N^{*}}$ should be replaced by 
$-m_{N^{*}}$.  Thus, the mass differences appear in the numerator of 
$\Pi_{1}^{\pi}$ term, and hence it vanishes for the diagonal 
couplings.  Therefore we try to derive the information of $\pi NN^{*}$ 
coupling from the $\Pi_{1}^{\pi}$ term of (~\ref{eq:cor}).

Next we see the theoretical side.  In recent reports, we have computed 
the two point correlation function (\ref{eq:cor}) in the operator 
product expansion (OPE) and found that it vanishes up to order 
dimension eight~\cite{jho}.  We have demonstrated that this result is 
a consequence of chiral symmetry of the interpolating field $J(x)$.  
First we discuss the transformation rule of the interpolating field 
(~\ref{eq:nucur}) under the $SU(2)$ axial transformation.  The quark 
fields, which belong to the fundamental representation of $SU(2)_{R} 
\times SU(2)_{L}$, are transformed by definition as
\begin{equation}
   [ Q_{5}^{a}, q ] = { 1 \over 2} i \gamma_{5} \tau^{a} q \ .
\end{equation}
Using this relation we obtain the commutation relation of the 
interpolating field (~\ref{eq:nucur}) as
\begin{equation}
   [ Q_{5}^{a}, J ] = { 1 \over 2} i \gamma_{5} \tau^{a} J \ , 
   \label{transJ}
\end{equation}
which is independent of the choice of $t$.  Note that this relation is 
not trivial because the interpolating field for the nucleon is made so 
as to satisfy the isospin symmetry of $SU(2)_{V}$, while axial 
properties are not subject to such a constraint.  Therefore the 
relation (~\ref{transJ}) is not the definition but is the consequence 
for the particular choice of the interpolating field 
(~\ref{eq:nucur}).

Now we rewrite the one pion matrix element (\ref{eq:cor}) in terms of 
the commutation relation with the axial charge $Q_{A}^{a}$:
\begin{eqnarray}
\Pi^{\pi^a} (p) &=& \lim_{q \to 0}  
\int d^4x e^{ipx} \bra{0} T J(x) \bar J(0) \ket{\pi^a(q)} 
\nonumber \\
&=& - \frac{i}{\sqrt{2}f_\pi} \int d^4x e^{ipx} \bra{0} [ Q_A^a , 
T J(x) \bar J(0) ] \ket{0} \nonumber \\
&=& - \frac{i}{2\sqrt{2}f_\pi} 
\int d^4x e^{ipx} 
\{ \gamma_5 \tau^a, \bra{0} T J(x) \bar J(0)  \ket{0} \}  \, .
\label{Q5comm}
\end{eqnarray}
Here we have used the transformation property (\ref{transJ}) of the 
interpolating field $J$.  Using the Lorentz structure of the vacuum to 
vacuum matrix element of (\ref{Q5comm}): 
\begin{equation}
   \int d^4 x e^{ipx} 
    \bra{0} J(x) \bar J(0) \ket{0} \sim A \pslash + B 1,
\end{equation}
the $\pslash \gamma_5$ term disappears in (\ref{Q5comm}).  This is the 
basis for the vanishing coupling constant for $g_{\pi NN^{*}}$.
The crucial point here is that the both 
interpolating field for $N$ and $N^{*}$ satisfy the same 
transformation rule under the $SU(2) \times SU(2)$ transformations.

We should make one remark here.  
When we write the phenomenological correlation function (~\ref{eq:pNN*}), 
only one term  for 
$N^* \to N + \pi$ has been considered, whereas in the 
theoretical expression (\ref{eq:cor}) 
another contribution from the reversed process 
$N + \pi \to N^*$  is also contained.  
If both the contributions are included, the phenomenological 
expression for the $\pslash \gamma_5$ term is factorized by 
$\lambda_{N}(t) \lambda_{N^{*}}(s) - 
 \lambda_{N}(s) \lambda_{N^{*}}(t)$.  
%Therefore, the $\pslash \gamma_5$ term vanishes  when $s = t$.   
On the other hand, in the OPE side, the $\pslash \gamma_5$ term 
has the common factor $s-t$~\cite{jho}.
Thus in both cases the correlation function vanishes trivially when
$s = t$.  
However, in Eq. (\ref{Q5comm}) the $\pslash \gamma_5$ term  
vanishes not by this trivial factor 
but due to the chiral symmetry of the interpolating field.   
This is what we would like to emphasize in this paper.  

The key of the above proof is that Eq.  (\ref{transJ}) is satisfied 
regardless the choice of $t$, as the 
nucleon interpolating field $J(x;t)$ transforms as the fundamental 
representation of the chiral group $SU(2)_{R} \times SU(2)_{L}$.  
To look at this point in some detail, let us investigate 
the algebraic structure of the interpolating field.  
The nucleon field which consists of three quarks belongs to 
the following irreducible representation of 
$SU(2)_L \times SU(2)_R$:
\begin{eqnarray}
\left[ \left( \textstyle{\hal{1}},0 \right) 
           + \left( 0, \textstyle{\hal{1}} \right) \right] ^{3}
&= &
\left[ \left( \textstyle{\hal{3}},0 \right) 
           + \left( 0, \textstyle{\hal{3}} \right) \right]
+ 3 \, \left[ \left( \textstyle{\hal{1}},1 \right) 
           + \left( 1, \textstyle{\hal{1}} \right) \right] \nonumber \\
& & 
\label{dectwo}
+ 3 \, \left[ \left( \textstyle{\hal{1}},\tilde 0 \right) 
           + \left( \tilde 0, \textstyle{\hal{1}} \right) \right]
+ 2 \, \left[ \left( \textstyle{\hal{\tilde 1}},0 \right) 
           + \left( 0, \textstyle{\hal{\tilde 1}} \right) \right] \, ,
\end{eqnarray}
where tildes imply that a pair of left or 
right quarks are coupled to the isospin singlet~\cite{coji}.  
The relevant terms for the nucleon are then 
$( \hal{1}, \tilde 0) + ( \tilde 0, \hal{1} )$ and 
$( \hal{\tilde 1}, 0) + ( 0, \hal{\tilde 1})$, which corresponds to 
two independent terms of the interpolating field, 
$J(x;t=-1)$ and $J(x;t=1)$, respectively.  
However, these two terms cannot be distinguished within 
$SU(2)_R \times SU(2)_L$, as they carry the same $SU(2)$ axial 
charge.  
This is the underlying reason that the commutation relation 
(\ref{transJ}) holds regardless the parameter $t$.  

Now we mention the $\eta N N^*$ coupling briefly.  
Since the $\eta$ meson is an isospin singlet,  
we investigate the $U(1)_A$ property of the nucleon.  
The nucleon field $J(x,t)$ transforms 
under $U(1)_A$ transformations as follows:  
\begin{eqnarray}
\ [ Q_A , J(x; t=-1) ] & = & \gamma_5 J(x; t=-1) \, , \\
\ [ Q_A , J(x; t=1)  ] & = & 3 \gamma_5 J(x; t=1) \, ,
\end{eqnarray}
where $Q_A$ is the $U(1)_A$ axial charge.  A crucial point here is 
that the transformation rule of $J$ differs for different values of 
$t$, that is, the interpolating field for $N^*$ transforms differently 
from $N$ under the $U(1)_A$ transformation.  Therefore the 
correlation function for $\eta$ cannot reduce to the anti-commutation 
relation with $\gamma_5$ unlike Eq.  (\ref{Q5comm}) for the pion.  
Thus the $g_{\eta NN^*}$ coupling need not vanish.

To summarize briefly, we have seen that symmetry properties of the 
interpolating field $J$ lead to $g_{\pi NN^{*}} = 0$ while $g_{\eta 
NN^{*}}$ does not need to vanish.  Phenomenologically, these 
properties seem to be well satisfied by the negative parity resonance 
$N(1535)$, suggesting that the properties of the resonance are 
strongly governed by chiral symmetry.

\section{Linear $\sigma$-models with $N_{+}$ and $N_{-}$}
In the previous section we have seen that the transformation rule of 
$J$ under the axial transformation causes the vanishing of $g_{\pi 
NN^{*}}$.  In this section, we consider sigma models with $N$ and 
$N^{*}$, which would reflect the nature of the chiral symmetry, and 
see how the above results are realized in these models.

First we define two nucleon fields, $N_{1}$ and $N_{2}$, which 
belong to a chiral multiplet $(\hal{1}, 0) \oplus (0, \hal{1})$.  In 
this definition, these fields transform under $SU(2)_{R} \times 
SU(2)_{L}$ naively as
\begin{eqnarray}
  N_{1R} \longrightarrow R N_{1R} & \hspace{1cm} & N_{1L} 
  \longrightarrow L N_{1L} \ ,\\
  N_{2R} \longrightarrow R N_{2R} & \hspace{1cm} & N_{2L} 
  \longrightarrow L N_{2L} \ ,
\end{eqnarray}
where $R$ ($L$) is the element of $SU(2)_{R}$ ($SU(2)_{L}$) and 
$N_{1R}$ ($N_{1L}$) is the right (left) component of the Dirac 
spinors, satisfying $\gamma_{5} N_{R} = N_{R}$, $\gamma_{5} N_{L} = - 
N_{L}$.  The parity of a baryon field $\psi$ can be chosen 
arbitrarily.  Here we assign that $N_{2}$ has negative parity.  (If it 
were assigned to have positive parity, the following argument would 
not change.)

Considering the transformation rule for the meson field under the 
chiral transformation $ M \equiv \sigma + i \vec{\tau} \cdot \vec{\pi} 
\rightarrow L M R^{\dagger}$, we construct the chiral invariant 
Lagrangian:
\begin{eqnarray}
   {\cal L}_{SU(2)} & = &\bar{N_1} \delslash N_1 + \bar{N_2} \delslash 
   N_2 + a \bar{N_1} (\sigma + i \gamma_5 \vec{\tau} \cdot \vec{\pi}) 
   N_1 + b \bar{N_2} (\sigma + i \gamma_5 \vec{\tau} \cdot \vec{\pi}) 
   N_2 + \nonumber \\
         & & c \{ \bar{N_1} (\gamma_5 \sigma + i \vec{\tau} \cdot 
         \vec{\pi}) N_2 - \bar{N_2} (\gamma_5 \sigma + i \vec{\tau} \cdot 
         \vec{\pi}) N_1 \} + {\cal L}_{M} \label{ordsu2lag}
\end{eqnarray}
where ${\cal L}_M$ is a meson part of the Lagrangian, which is not 
important in this discussion. The fifth term in this Lagrangian 
gives a mixing of $N_{1}$ and $N_{2}$, which is also chiral invariant. 
We calculate 
the axial charge from this Lagrangian and the commutation relations of 
the axial charge and the nucleon fields.  We find the same commutation 
relations for $N_{1}$ and $N_{2}$:
\begin{equation}
   [Q_{5}^{a}, N_{1}] =  \hal{1} i\gamma_{5} \tau^{a} N_{1} \hspace{1cm}
   \ [Q_{5}^{a}, N_{2}] =  \hal{1} i\gamma_{5} \tau^{a} N_{2} 
\end{equation}

The chiral symmetry breaks down spontaneously with a finite vacuum 
expectation value of $\sigma$. To obtain physical nucleons $N_{+}$ 
and $N_{-}$ in the chiral broken phase, we have to diagonalize the 
mass matrix $M$, which is given by
\begin{equation}
  M \sim \sigma_{0} \left(
      \begin{array}{cc}
        a & \gamma_{5} c  \\
        - \gamma_{5} c & b
      \end{array}
    \right)
\end{equation}
where $\sigma_{0} \equiv \bra{0} \sigma \ket{0}$. This matrix can be 
diagonalized by defining the physical nucleon fields as 
\begin{equation}
   \left( 
       \begin{array}{c}
        N_{+}  \\
        N_{-}
       \end{array}
   \right) = {1 \over \sqrt{2 \cosh \delta }} 
        \left(
             \begin{array}{cc}
             -e^{-\delta/2}      & \gamma_{5} e^{\delta/2}  \\
                \gamma_{5} e^{\delta/2} & e^{-\delta/2}
             \end{array}
        \right)
        \left(
             \begin{array}{c}
                N_{1}  \\
                N_{2}
             \end{array}
        \right) \label{N+N-}
\end{equation}
where the mixing angle $\delta$ is defined as $\sinh \delta = {a + b 
\over 2 c}$.  The masses of $N_{+}$ and $N_{-}$ are given by
$  m_{\pm} = \sigma_{0} / 2 ( \sqrt{(a + b)^{2} + 4c^{2}} 
   \mp (a-b) ) $.

After the diagonalization the physical nucleon fields $N_{+}$ and $N_{-}$ 
decouple from each other because the coupling (matrix $C$) with pions 
takes the same form as the mass matrix $M$:
\begin{equation}
    C \sim \left(
      \begin{array}{cc}
        a & \gamma_{5} c  \\
        - \gamma_{5} c & b
      \end{array}
    \right) \ ,
\end{equation}
and can be diagonalized simultaneously with $M$.  Therefore the 
$g_{\pi N_{+}N_{-}}$ vanishes.  This means that this sigma model 
is just the sum of two sigma models for two independent nucleons.
In fact, $N_{+}$ and $N_{-}$ are independent even in the chiral 
restored phase, as the Lagrangian (~\ref{ordsu2lag}) can be 
diagonalized by the same matrix (~\ref{N+N-}) and becomes two 
independent sigma models. This is so because the mixing angle $\delta$ 
is independent of $\sigma_{0}$.
% We can diagonalized the mixing term between $N_1$ and $N_2$ in  
% the Lagrangian (~\ref{ordsu2lag}) redefining $N_1$ and $N_2$ to $N_+$ and
% $N_-$ with (~\ref{N+N-}) in the stage of constructing Lagrangian 
% (~\ref{ordsu2lag}). This redefinition of the nucleon fields has 
% nothing to do with the chiral symmetry breaking.

Another assignment of the chiral transformation for nucleons is 
possible.  The second nucleon $\psi_{2}$ is allowed to transform under 
the chiral symmetry in the reversed way to the first nucleon 
$\psi_{1}$, that is,
\begin{eqnarray}
   \psi_{1R} \longrightarrow R \psi_{1R} & \hspace{1cm} &  \psi_{1L} 
   \longrightarrow L \psi_{1L} \\
   \psi_{2R} \longrightarrow L \psi_{2R} & \hspace{1cm} &  \psi_{2L} 
   \longrightarrow R \psi_{2L} \ .
\end{eqnarray}
$\psi_{2}$, which is defined is called ``mirror fermion''.  Note that 
the right component of $\psi_2$ transforms under the ``left'' chiral 
transformation and the left transforms under the ``right'' chiral 
transformation.  In the language of the representation of the $SU(2)_R 
\times SU(2)_L$, $\psi_{2R}$ and $\psi_{2L}$ belong to the irreducible 
representation $(0, \hal{1})$ and $(\hal{1}, 0)$ respectively, while 
$\psi_{1R}$ and $\psi_{1L}$ belong to $(\hal{1}, 0)$ and $(0,\hal{1})$ 
respectively.  The reason that this assignment is possible is that the 
left-right of fermion is independent from the left-right of the chiral 
symmetry, although we use the same word, because the chirality of 
fermion is the representation of Lorentz group while the chirality of 
the chiral symmetry is the one of the chiral group.  We can not allow 
these assignment for quarks by definition.  However we can allow such 
mirror assignment for baryons, because baryons are composite of quarks 
and gluons.  Although the parity of $\psi_{2}$ is arbitrary, we here 
define that $\psi_{2}$ represents a negative parity baryon.

The chirally invariant Lagrangian, which was proposed and studied first by 
DeTar and Kunihiro~\cite{dk}, is then
\begin{eqnarray}
     {\cal L} & = & \bar{\psi_1} i \delslash \psi_1 + 
           \bar{\psi_2} i \delslash \psi_2 +
            m_{0}( \bar{\psi_2} \gamma_{5} \psi_1 - \bar{\psi_1} 
            \gamma_{5} \psi_2  ) 
                \nonumber \\ 
      & &    + a \bar{\psi_{1}} (\sigma + i \gamma_5 \vec{\tau} \cdot 
           \vec{\pi}) \psi_{1} + 
        b \bar{\psi_{2}} (\sigma + i \gamma_5 \vec{\tau} \cdot 
        \vec{\pi}) \psi_{2} + {\cal L}_{M}.
  \label{mirsu2lag}
\end{eqnarray}
This Lagrangian has the chirally invariant mass term $ m_{0}( 
\bar{\psi_2} \gamma_{5} \psi_1 - \bar{\psi_1} \gamma_{5} \psi_2 ) $, 
and therefore the nucleons are massive even in the chiral restored 
phase.  Due to this term this Lagrangian is no longer the sum of the 
two independent sigma models, since the $\psi_{1R}$ and $\psi_{2L}$ 
(or $\psi_{1L}$ and $\psi_{2R}$), which are in the same multiplet, are 
related by the chiral invariant mass term.  Thus when we write the nucleon 
fields on the physical basis to diagonalize the mass term, the coupling 
between positive and negative parity nucleons with pions need not  
vanish.  (In the chiral restored phase the diagonal coupling rather 
vanishes !)

A phenomenological significance is that axial charges of the nucleons 
in this model have the opposite sign to each other.  In the 
spontaneous symmetry breaking phase the mass matrix is diagonalized by 
$\psi_{+}$ and $\psi_{-}$, which are defined as
\begin{equation}
  \left( \begin{array}{c} \psi_+ \\ \psi_- \end{array} \right) = 
  \frac{1}{\sqrt{2 \cosh \delta}} \left(
             \begin{array}{cc}
             -e^{-\delta/2}      & \gamma_{5} e^{\delta/2}  \\
                \gamma_{5} e^{\delta/2} & e^{-\delta/2}
             \end{array}
        \right)
    \left( \begin{array}{c} \psi_{1} \\ \psi_{2} \end{array} \right)
\end{equation}
with $\sinh \delta = (a + b) \sigma_0 / 2 m_0$.
In this basis $(\psi_{+}, \psi_{-})$, the axial-vector charges $g_A$ are
\begin{equation}
  g_A  = \left( \begin{array}{cc}
    \tanh \delta            & -\frac{1}{\cosh \delta} \\
    -\frac{1}{\cosh \delta} & -\tanh \delta
  \end{array} \right) \ .
\end{equation}
Thus the sign of $g_{A\psi_{-}\psi_{-}}$ is opposite to 
$g_{A\psi_{+}\psi_{+}}$.  It would be interesting if we could 
distinguish which type of fermion (naive or mirror) is realized in 
observed $N^{*}$ by measuring $g_{A}$ of excited nucleons.

Finally we comment on these $\sigma$-models from the viewpoint of the 
chiral partner.  Particles in a chiral partner can be transformed each 
other under chiral transformation and therefore are degenerate in the 
chiral restored phase.  For example $\sigma$ and $\pi$ are chiral 
partners and belong to the $(\hal{1},\hal{1})$ representation.  In the 
following argument we only consider the chiral symmetric phase because 
the chiral partner becomes degenerate in this phase.  The chiral partner of a 
fermion $\psi$ is defined as
\begin{equation}
   \psi^\prime \equiv [Q_{5}^{a} , \psi] \ ,
\end{equation}
similarly with $i \pi^{a} = [ Q_{5}^{a} , \sigma] $.  

There are two cases.  First is the case that the fermions are massless 
in the chiral restored phase.  In this case it is sufficient to have 
only one kind $N_{+}$ of fermions.  Calculating the commutation 
relation of the axial charge $Q_{5}^{a}$ and $N_{+}$, we find
\begin{eqnarray}
   N_{+}^\prime & = & [Q_{5}^{a}, N_+ ] \\
               & = & \gamma_5 \tau^a N_+ \ .
\end{eqnarray}
This means that the chiral partner of $N_{+} \equiv N_{+R} + N_{+L}$ is 
$N_{+R} - N_{+L}$.  The second case is that the fermion is massive 
even in the chiral restored phase.  In this case we need to introduce 
two kinds of fermions $\psi_{1}$ and $\psi_{2}$ in the mirror 
construction.  Similarly calculating $Q_{5}^{a}$ from the chiral 
invariant Lagrangian (~\ref{mirsu2lag}), we obtain
\begin{eqnarray}
   \psi_{+}^{\prime} & = & [Q_{5}^{a} , \psi_{+} ] \\
             & = & \tau^{a} \psi_{-} \ ,
\end{eqnarray}
where $\psi_{+}$ and $\psi_{-}$ are given from $\psi_{1}$ and 
$\psi_{2}$ by diagonalizing the mass term in (~\ref{mirsu2lag}).  In 
this case the chiral partner of $\psi_{+}$ is $\psi_{-}$, which is 
closer to the original meaning of chiral partner than the first case.

Our QCD sum rule approach belongs to the first case.  $N^*$ in our QCD 
sum rule transforms in the same way as $N$ under $SU(2) \times SU(2)$ 
chiral transformation because of its construction given in eqs.\ 
(~\ref{constN}) and (~\ref{constN*}).  Therefore the  suppression of the $\pi 
NN^*$ coupling is caused by the chiral assignment for $N^*$.  In the 
second case, however, DeTar and Kunihiro  reproduced 
the observed quantities choosing $m_0 = 270$ MeV \cite{dk}.  
Moreover we extended this 
model to $SU(3)$ and analyzed the decays of $\hal{1}^-$ baryons in the 
strong interaction using the chiral perturbation theory~\cite{nemoto}.  
Both the calculations suggest that the suppression of the $\pi NN^*$ 
coupling is caused by the smallness of $m_0$, which is one of the 
parameters determining the mixing angle $\delta$.  
In order to determine which of the cases is realized in the physical $N^*$,
we need to investigate more on the chiral structure of the 
excited nucleon, such as the sign of $g_A$.

\section{Conclusion}
We have investigated the $\pi NN^{*}$ coupling based on the QCD sum 
rule in the soft and chiral limit.  We have seen that the coupling 
vanishes due to an algebraic property of the interpolating field.  It 
is essential that the interpolating fields for $N$ and $N^{*}$ have 
the same chiral property.  This suppression is consistent with the 
smallness of the observed $g_{\pi NN^{*}}$ compared with $g_{\pi NN}$.  
Similar discussions on $g_{\eta NN^{*}}$ conclude that the $g_{\eta 
NN^{*}}$ need not vanish because the interpolating field for $N$ and 
$N^{*}$ transform under $U(1)_{A}$ in different ways.

To clarify the relation between vanishing $g_{\pi NN^{*}}$ and chiral 
symmetry, we have constructed linear sigma models.  We have two ways 
for assignments of the chiral transformation to the negative parity 
nucleon.  The first case is when the transformation rule for $N^{*}$ 
is the same as one for $N$.  In this assignment $N$ and $N^{*}$ 
transform in the same way under the axial transformation and cannot 
have masses in the chiral restored phase.  The nucleons decouple each 
other because this model is only the sum of two independent sigma 
models.  Our QCD sum rule approach belongs to this case.  The second 
is when $N^{*}$ transforms in the reversed way to $N$, namely, $N_{L}^{*}$ 
($N_{R}^{*}$) transforms under the $SU(2)_{R}$ ($SU(2)_{L}$) 
transformation.  In the sigma model with this assignment nucleons are 
massive even in the chiral restored phase.  $N$ and $N^{*}$ have the 
opposite sign of axial charges each other in the spontaneously broken 
phase.  And the coupling of nucleons does not need to vanish.

We have seen that the $\pi NN^{*}$ coupling is strongly related to the 
chiral symmetry of $N$ and $N^{*}$.  There are two types of fermions, 
and it is interesting to see which assignment are realized in the 
physical excited nucleons.  Therefore we need to investigate more the 
chiral structure of baryons to understand the properties of baryons.

\end{document}